\documentclass[twocolumn,trackchanges]{aastex631}
\shorttitle{ALMA [\ion{O}{3}] and [\ion{C}{2}] Detections of A1689-zD1 at $z=7.13$}
\shortauthors{Wong et al.}
\graphicspath{{./}{figures/}}

\begin{document}

\title{ALMA detections of [\ion{O}{3}] and [\ion{C}{2}] emission lines from A1689-zD1 at $z=7.13$}

\correspondingauthor{Yi Hang Valerie Wong}
\email{valeriew510@gmail.com}

\author[0000-0001-6236-6882]{Yi Hang Valerie Wong}
\affiliation{Institute of Astronomy, National Tsing Hua University, 101, Section 2. Kuang-Fu Road, Hsinchu, 30013, Taiwan (R.O.C.)}

\author[0000-0003-2658-5871]{Poya Wang}
\affiliation{Department of Physics, Tamkang University, New Taipei City 251301, Taiwan (R.O.C.)}

\author[0000-0001-7228-1428]{Tetsuya Hashimoto}
\affiliation{Institute of Astronomy, National Tsing Hua University, 101, Section 2. Kuang-Fu Road, Hsinchu, 30013, Taiwan (R.O.C.)}
\affiliation{Centre for Informatics and Computation in Astronomy (CICA), National Tsing Hua University, 101, Section 2. Kuang-Fu Road, Hsinchu, 30013, Taiwan (R.O.C.)}
\affiliation{Department of Physics, National Chung Hsing University, No. 145, Xingda Rd., South Dist., Taichung, 40227, Taiwan (R.O.C.)}

\author{Toshinobu Takagi}
\affiliation{Japan Space Forum, Shin-Ochanomizu Urban Trinity Bldg. 3F, 3-2-1, Kandasurugadai, Chiyoda-ku, Tokyo 101-0062 Japan}

\author[0000-0002-6821-8669]{Tomotsugu Goto}
\affiliation{Institute of Astronomy, National Tsing Hua University, 101, Section 2. Kuang-Fu Road, Hsinchu, 30013, Taiwan (R.O.C.)}
\affiliation{Department of Physics, National Tsing Hua University, 101, Section 2. Kuang-Fu Road, Hsinchu, 30013, Taiwan (R.O.C.)}

\author[0000-0001-9970-8145]{Seong Jin Kim}
\affiliation{Institute of Astronomy, National Tsing Hua University, 101, Section 2. Kuang-Fu Road, Hsinchu, 30013, Taiwan (R.O.C.)}


\author{Cossas K.-W. Wu}
\affiliation{Department of Physics, National Tsing Hua University, 101, Section 2. Kuang-Fu Road, Hsinchu, 30013, Taiwan (R.O.C.)}

\author[0000-0003-4479-4415]{Alvina Y. L. On}
\affiliation{Institute of Astronomy, National Tsing Hua University, 101, Section 2. Kuang-Fu Road, Hsinchu, 30013, Taiwan (R.O.C.)}
\affiliation{Centre for Informatics and Computation in Astronomy (CICA), National Tsing Hua University, 101, Section 2. Kuang-Fu Road, Hsinchu, 30013, Taiwan (R.O.C.)}
\affiliation{Mullard Space Science Laboratory, University College London, Holmbury St. Mary, Dorking, Surrey RH5 6NT, UK}

\author[0000-0002-5687-0609]{Daryl Joe D. Santos}
\affiliation{Institute of Astronomy, National Tsing Hua University, 101, Section 2. Kuang-Fu Road, Hsinchu, 30013, Taiwan (R.O.C.)}
\affiliation{Max Planck Institute for Extraterrestrial Physics, Gie{\ss}enbachstra{\ss}e 1, 85748 Garching, Germany}

\author[0000-0002-4965-6524]{Ting-Yi Lu}
\affiliation{Institute of Astronomy, National Tsing Hua University, 101, Section 2. Kuang-Fu Road, Hsinchu, 30013, Taiwan (R.O.C.)}
\affiliation{Cosmic Dawn Center, Niels Bohr Institute, University of Copenhagen, Jagtvej 128
2200 Copenhagen N, Denmark}

\author[0000-0002-9119-2313]{Ece Kilerci-Eser}
\affiliation{Sabanc{\i} University, Faculty of Engineering and Natural Sciences, 34956, Istanbul, Turkey}

\author[0000-0002-8560-3497]{Simon C.-C. Ho}
\affiliation{Institute of Astronomy, National Tsing Hua University, 101, Section 2. Kuang-Fu Road, Hsinchu, 30013, Taiwan (R.O.C.)}

\author{Tiger Y.-Y. Hsiao}
\affiliation{Institute of Astronomy, National Tsing Hua University, 101, Section 2. Kuang-Fu Road, Hsinchu, 30013, Taiwan (R.O.C.)}



\begin{abstract}

A1689-zD1 is one of the most distant galaxies, discovered with the aid of gravitational lensing, providing us with an important opportunity to study galaxy formation in the very early Universe.
In this study, we report the detection of [\ion{C}{2}]158\,{\rm{$\mu$m}} and [\ion{O}{3}]88\,{\rm{$\mu$m}} emission lines of A1689-zD1 in the ALMA Bands 6 and 8.
We measure the redshift of this galaxy as $z_{\rm{sys}}=7.133\pm0.005$ based on the [\ion{C}{2}] and [\ion{O}{3}] emission lines, 
consistent with that adopted by \citet{2021MNRAS.tmpL..91B}.
The observed $L_{[\rm{O\,III]}}/L_{[\rm{C\,II]}}$ ratio is $2.09\pm0.09$, higher than most of the local galaxies, but consistent with other $z\sim7$ galaxies.
The 
moderate-spatial resolution of ALMA data provided us with a precious opportunity to investigate spatial variation of  $L_{[\rm{O\,III]}}/L_{[\rm{C\,II]}}$.
In contrast to the average value of 2.09, 
we find a much higher $L_{[\rm{O\,III]}}/L_{[\rm{C\,II]}}$ of $\sim 7$ at the center of the galaxy.
This spatial variation of $L_{[\rm{O\,III]}}/L_{[\rm{C\,II]}}$ was seldom reported for other high-z galaxies. It is also interesting that the peak of the ratio does not overlap with optical peaks.
Possible physical reasons include a central AGN, shock heating from merging, and starburst.
Our 
moderate-spatial resolution data also reveals 
that in addition to the observed two clumps shown in previous HST images, there is a redshifted segment to the west of the northern optical clump. Such a structure is consistent with previous claims that A1689-zD1 is a merging galaxy, but with the northern redshifted part being some ejected materials, or that the northern redshifted materials being from a third more highly obscured region of the galaxy. 

\end{abstract}



\section{Introduction}\label{sec:intro}

Studies of high-redshift (high-$z$) galaxies are crucial in understanding the early phase of galaxy formation and evolution. 
A gravitationally-lensed galaxy A1689-zD1 is among the most distant sources ($z \sim$ 7.13) discovered so far, strongly lensed by a factor of $\mu=$9.3 \protect\citep{2008ApJ...678..647B}.
It is, therefore, a useful probe to study the early Universe.
Recently in \citet{2021MNRAS.tmpL..91B}, they adopted a new redshift of the galaxy $z=$7.13
, and our measurement confirms the redshift.

Having a high dust mass as compared to the Milky Way makes A1689-zD1 unusual among high-redshift dust emitters. It is also a sub-L* galaxy \citep{2015Natur.519..327W}. The massive amount of dust may indicate a gas-rich system that makes the emission lines from interstellar medium (ISM) detectable. 
According to \protect\citet{2008ApJ...678..647B}, A1689-zD1 is a system composed of two clumps, which are likely separate star-forming regions within the galaxy, but they could conceivably be interpreted as small star-forming galaxies merging at high redshift.


Previously, A1689-zD1 was detected in dust continuum with the Atacama Large Millimeter/submillimeter Array (ALMA) in the absence of emission lines \protect\citep{2017MNRAS.466..138K, 2021MNRAS.tmpL..91B}, except for a slight excess in [\ion{C}{2}]158\,{\rm{$\mu$m}} \protect\citep{2017MNRAS.466..138K}. Using the dust continuum, the sizes of
the two clumps detected were estimated to be $0.4-1.7$\,kpc, and the dust temperature was approximated to be $T_{\rm{dust}} \sim 35-45$\,K. 
A recent update of dust temperature and mass of A1689-zD1 using continuum estimated from [\ion{C}{2}]-based method are $T_{\rm{dust}}=40^{+13}_{-7}$K and $M_{\rm{d}}=2.0^{+1.8}_{-1.0}\times10^{7}$M$_\odot$ \citep{2021MNRAS.tmpL..91B}.

In \protect\citet{2015Natur.519..327W}, the redshift was measured from the X-shooter spectra taken on the Very Large Telescope (VLT), using the Ly$\alpha$ break due to the lack of Ly$\alpha$ and other emission lines.
However, the redshift measurement based on the Ly$\alpha$ break is less certain than that using high excitation lines.
Moreover, the break is close to the spectrograph’s near-infrared (NIR) / visual (VIS) arm split. Thus, the redshift needs to be measured more accurately. Using [\ion{C}{2}] and [\ion{O}{3}] 
emission lines, a redshift of 7.13 by spectral measurement adopted by \citet{2021MNRAS.tmpL..91B} is confirmed in this work.

Since [\ion{O}{3}]88\,{\rm{$\mu$m}} is bright for some galaxies at $z>6$ \citep[e.g.,][]{2016Sci...352.1559I,2017A&A...605A..42C,2020ApJ...896...93H}, the [\ion{O}{3}] line is one of the most useful tracers for ISM properties. In addition, [\ion{O}{3}] has a higher ionization potential (35.1 eV) than [\ion{C}{2}] (11.3 eV). Therefore, the [\ion{O}{3}]/[\ion{C}{2}] ratio is useful to investigate the ionization state.

In this paper, we present an analysis of archival ALMA data of [\ion{C}{2}]158\,{\rm{$\mu$m}} and [\ion{O}{3}]88\,{\rm{$\mu$m}} in Cycles 3 and 5 with a higher spatial resolution ($\sim$0$^{\prime\prime}$.2) than ever before ($\sim$0$^{\prime\prime}$.9 \citep{2017MNRAS.466..138K}; 0$^{\prime\prime}$.6-0$^{\prime\prime}$.7 \citep{2015Natur.519..327W}).
With this higher spatial resolution, the data can be used to examine the line emissions, resolve structures inside the galaxy, investigate velocity structures, etc., in comparison with the HST images.
For example, 
this higher-spatial resolution allows us to 
examine the kinematic properties of A1689-zD1.
We can also investigate the ratio of [\ion{O}{3}]/[\ion{C}{2}] as a function of positions within the galaxy.
This is important because, as some previous studies discussed, the observed high [\ion{O}{3}]/[\ion{C}{2}] may be due to the difference in their spatial distributions \citep[]{2017A&A...605A..42C}.

This paper is organized as follows. In section~\ref{sec:observe} we describe the ALMA dataset used in the study; section~\ref{sec:results} presents the main results, with discussions in section~\ref{sec:discussion}. Finally, we present our conclusions in section~\ref{sec:conclusions}. Throughout this paper, we assume the {\it Planck15} cosmology \citep{2016A&A...594A..13P} as a fiducial model, i.e., a $\Lambda$ cold dark matter cosmology with ($\Omega_{m}$,$\Omega_{\Lambda}$,$\Omega_{b}$,$h$)=(0.308, 0.682, 0.0486, 0.678).

\section{Observation}\label{sec:observe}

Two ALMA bands, Bands 6 and 8, are chosen in this study in order to characterize the physical properties of A1689-zD1.
Because it is a dusty galaxy \citep[]{2015Natur.519..327W}, it is important to measure the star-formation rate (SFR) derived from far-infrared (FIR) observations, using its relation with [\ion{C}{2}]158\,{\rm{$\mu$m}} and [\ion{O}{3}]88\,{\rm{$\mu$m}} emissions.
At the redshift of A1689-zD1, Band 6 covers [\ion{C}{2}], and Band 8 covers [\ion{O}{3}] emission lines.


The Band 6 observation was carried out in 2016 from August 1 to August 25 (Cycle 3) with the ALMA 12m array, and an antenna configuration of C36-(4)/5 and C36-(5)/6 (project ID: 2015.1.01406.S, P.I.: D. Watson);
while the Band 8 data was obtained from observations in 2018 from April to December (Cycle 5), also with 12m array, but an antenna configuration of C43-3 and C43-4 (project ID: 2017.1.00775.S, P.I.: D. Watson).
The observational configurations are summarized in Table~\ref{Tab:Observational configurations}.

The calibration of visibility data from ALMA archive was conducted with the pipeline versions 4.7 and 5.4, for Band 6 and 8 respectively, provided by the ALMA project using the Common Astronomy Software Applications (CASA).
For the Band 6 data targeting [\ion{C}{2}], the channel width of the observation is 1.95 MHz. As for the Band 8 data targeting [\ion{O}{3}], the channel width is 3.91 MHz.
The total exposure time on source for Band 6 ([\ion{C}{2}]) observation is 28,123 seconds $\sim$ 7.81 hours,
while that for Band 8 ([\ion{O}{3}]) observation is 12,337 seconds $\sim$ 3.43 hours.
The maximum recoverable scale (MRS) for the Band 6 and Band 8 data are 5$^{\prime\prime}$.32 and 4$^{\prime\prime}$.58, which are larger than the detected signals as we discuss below.

\begin{deluxetable*}{lcc}[ht!]
\tablenum{1}
\tablecaption{Observational configurations\label{Tab:Observational configurations}}
\tablewidth{0pt}
\tablehead{
\colhead{} & \colhead{Band 6} & \colhead{Band 8} \\
\colhead{} & \colhead{[\ion{C}{2}]158\,{\rm{$\mu$m}}} & \colhead{[\ion{O}{3}]88\,{\rm{$\mu$m}}}
}
\startdata
Baseline lengths (m) & 15 - 1605 & 15 - 784 \\
Number of antennas & 38 - 40 & 41 - 47 \\
Spectral window 1 (GHz) & 217.17 - 219.05 & 404.48 - 406.30 \\
Spectral window 2 (GHz) & 219.05 - 220.93 & 406.31 - 408.14 \\
Spectral window 3 (GHz) & 231.96 - 233.84 & 416.33 - 418.20 \\
Spectral window 4 (GHz) & 233.85 - 235.73 & 418.18 - 420.01 \\
Channel width (MHz)%
& 1.95 & 3.91 \\
Exp. time on source (hour) & 7.81 & 3.43 \\
PWV (mm) & 0.3 - 1.1 & 0.3 - 1.0 \\
MRS (arcsec) & 5.32 & 4.58 \\
\enddata
\tablecomments{The PWV values for each of the band data are from separate measurement sets due to the long exposure time; therefore, we present the minimum and maximum of averages per Execution Block (EB).}
\end{deluxetable*}

Using the calibrated measurement set (MS), we create image cubes for [\ion{C}{2}] and [\ion{O}{3}] lines with a spectral resolution of 10\,km\,s$^{-1}$.
The beam size is controlled by the robustness parameter of the Briggs weighting and $uv$-tapering in the \texttt{CASA} task \texttt{tclean()}.
First, we subtract the continuum in each band using the \texttt{CASA} task \texttt{uvcontsub} by selecting the frequency ranges without line detection, i.e., 232.000$\sim$233.420 GHz for Band 6; 416.342$\sim$416.823GHz and 417.588$\sim$418.178 GHz for Band 8.
Next, in order to investigate both the morphology and photometry of A1689-zD1, we produce 2 sets of image cubes for each of the bands \citep{2019PASJ...71...71H}.
Since we are interested in the extent 
of [\ion{C}{2}] and [\ion{O}{3}] emissions of the galaxy, we adopt a Briggs robust weighting of R$=$0.5, achieving a beam size of $0^{\prime\prime}.271\times0^{\prime\prime}.243$ ($0^{\prime\prime}.387\times0^{\prime\prime}.332$), with a position angle at 
$-$80$^{\circ}$ ($-$89$^{\circ}$) for Band 6 (8) data.
As presented in section~\ref{subsec:spatial distribution}, the extended structure does not exceed the MRS in both bands.
Hence, we use the image cube sets with Briggs robust parameter R$=$0.5 for morphology analysis of A1689-zD1.
As for maximizing the sensitivity of collecting the flux as a point source, we use natural weighting by setting R$=$2, and further apply $uv$-tapering to 2$^{\prime\prime}$ so as to get the maximum flux of the extended source.
The beam size is dramatically increased to $2^{\prime\prime}.640\times2^{\prime\prime}.337$ ($1^{\prime\prime}.874\times1^{\prime\prime}.790$) at
$-$63$^{\circ}$ ($-$70$^{\circ}$) for Band 6 (8) data.
This set of data is used for flux and related calculations of the galaxy.
The characteristics of the calibrated images are summarized in Table~\ref{Tab:image characteristics}.

\begin{deluxetable*}{lcccc}[ht!]
\tablenum{2}
\tablecaption{Summary of image characteristics\label{Tab:image characteristics}}
\tablewidth{0pt}
\tablehead{
\colhead{} & \colhead{Band 6} & \colhead{Band 6} & \colhead{Band 8} & \colhead{Band 8} \\
\colhead{} & \colhead{(Briggs)} & \colhead{(natural \& $uv$-tapered)} & \colhead{(Briggs)} & \colhead{(natural \& $uv$-tapered)}
}
\startdata
Beam major / minor axes (arcsec) & 0.271 / 0.243 & 2.640 / 2.337 & 0.387 / 0.332 & 1.874 / 1.790 \\
Beam position angle$^{a}$ (deg) & -80 & -63 & 89 & -70 \\
Sensitivity$^{b}$ (mJy beam$^{-1}$)
& 0.17 & 0.59 & 0.66 & 1.69 \\
\enddata
\tablecomments{$^{a}$The beam PA is in degrees east of north by which the major axis is rotated. $^{b}$Sensitivity is defined as the RMS value of the cube image with 10\,km\,s$^{-1}$ channel width.}
\end{deluxetable*}

\section{Results}\label{sec:results}
\subsection{[\texorpdfstring{\ion{C}{2}}{CII}] and [\texorpdfstring{\ion{O}{3}}{OIII}] emission lines}\label{sec:emission lines}

\subsubsection{Spatial distribution}\label{subsec:spatial distribution}

\begin{figure}[ht!]
    \epsscale{1.15}
    \plotone{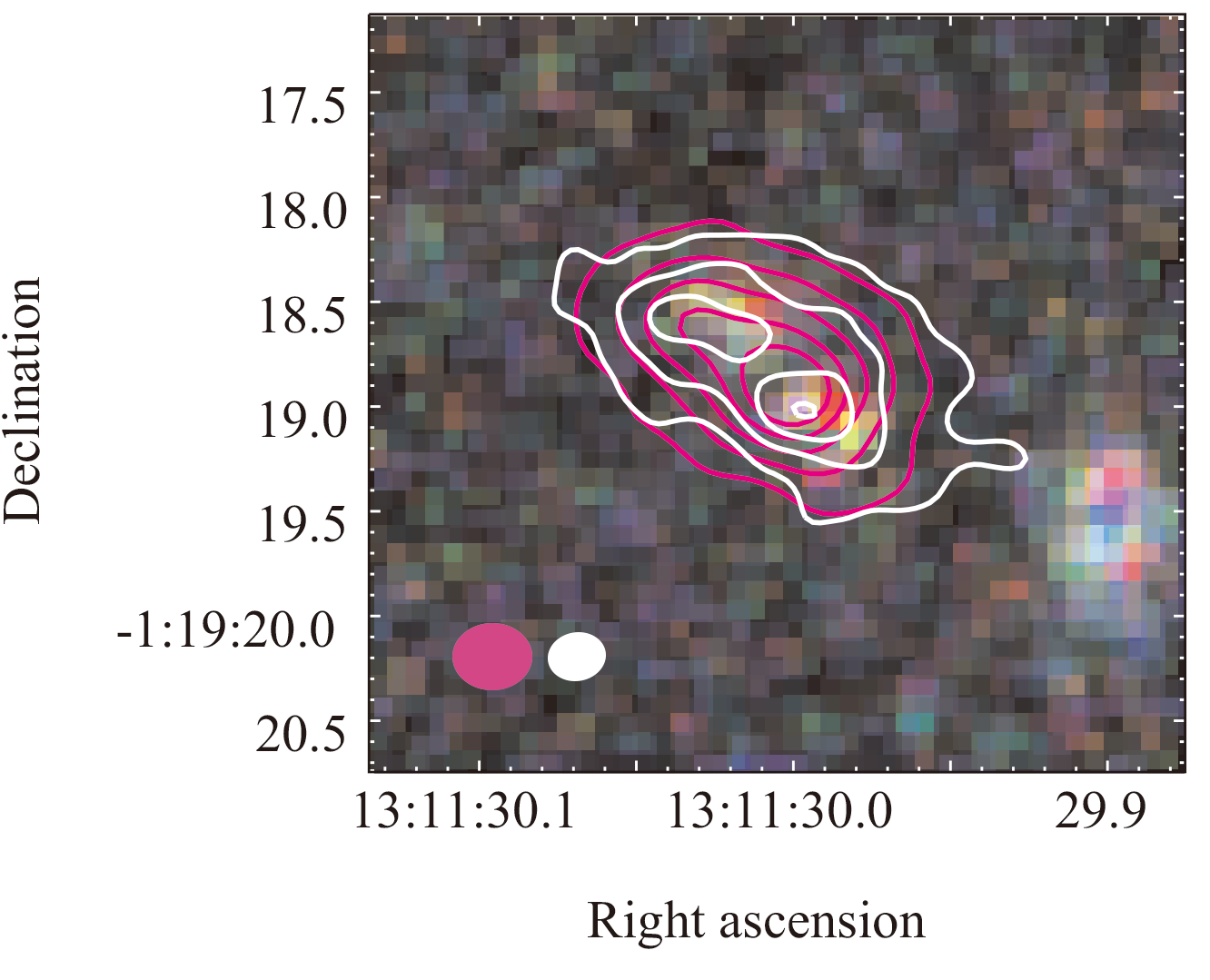}
    \caption{[\ion{C}{2}] (white) contours are drawn at 3, 7, 11, and 15 $\sigma$ levels ($\sigma_{\rm{[C\,II]}}\sim17.1\, \rm{mJy\,beam^{-1}\,km\,s^{-1}}$), while [\ion{O}{3}] (magenta) contours are drawn at 3, 7, 11, 15, and 19 $\sigma$ levels ($\sigma_{\rm{[O\,III]}}\sim55.7\,\rm{mJy\,beam^{-1}\,km\,s^{-1}}$). Overlaid is a $\sim 3^{\prime\prime}.5\times3^{\prime\prime}.5$ cutout image of HST data composed of images using the filters: F105W(blue), F125W(green), and F160W(red). The corresponding beam sizes of the Briggs Band 6 and Band 8 images used are $0^{\prime\prime}.271\times0^{\prime\prime}.243$ at $-80^{\circ}$ and $0^{\prime\prime}.387\times0^{\prime\prime}.332$ at $89^{\circ}$, respectively. 
    }
\label{Fig:mom0}
\end{figure}

Using \texttt{CASA} task \texttt{immoments}, we obtain moment 0 maps of the Band 6 and Band 8 (Briggs) data by stacking the cube channels along the spectrum, covering $\sim$ 400 km s$^{-1}$ ($\sim$ 1.2 FWHM).
It is to make sure the detections of emission lines are valid \citep{2019ApJ...881...63N}.
This range of frequency is comparable to that in previous studies \citep{2016Sci...352.1559I,2017ApJ...851..145M,2019PASJ...71..109H,2019ApJ...881...63N}.
Fig.~\ref{Fig:mom0} shows the [\ion{C}{2}] (white contours) line emission at 3, 7, 11, and 15$\sigma$ levels, and [\ion{O}{3}] (magenta contours) line emission at 3, 7, 11, 15, and 19$\sigma$ levels, overlaid on the combined Hubble Space Telescope (HST) color image.
For the purpose of color display, we stack HST filters with F105W as blue color; F125W as green; and F160W as red (proposal ID: 11802, P.I.: Ford).
There is a known coordinate offset between their ALMA image and HST image as an average value of 0$^{\prime\prime}.4\sim0^{\prime\prime}$.45 \citep{2015Natur.519..327W}.
We, thus, measure the position difference of the peaks between the ALMA Bands 6 and 8 data and the HST image we use, and shift the ALMA contours by 0$^{\prime\prime}$.386 along right ascension (RA) in order to match the peaks for a better comparison.
Because the structures shown in Fig.~\ref{Fig:mom0} extend to a 3$\sigma$ detection that is within an area of $\sim 2^{\prime\prime}.3\times1^{\prime\prime}.5$ for both [\ion{C}{2}] and [\ion{O}{3}],
which is much smaller than the MRS of 5$^{\prime\prime}$.32 and 4$^{\prime\prime}$.58 for [\ion{C}{2}] and [\ion{O}{3}], respectively,
we confirm that the structures observed are not artificial.
Both [\ion{C}{2}] and [\ion{O}{3}] emission lines are spatially well resolved with angular resolutions of 0$^{\prime\prime}$.203 and 0$^{\prime\prime}$.232, respectively.

From the HST image shown in Fig.~\ref{Fig:mom0}, we see that there are two clumps: one on the northeast (clump A), and one on the southwest (clump B).
We refer them as clump A and clump B throughout the context.


\subsubsection{Spectra}\label{subsec:spectra}

\begin{figure*}
    \gridline{\fig{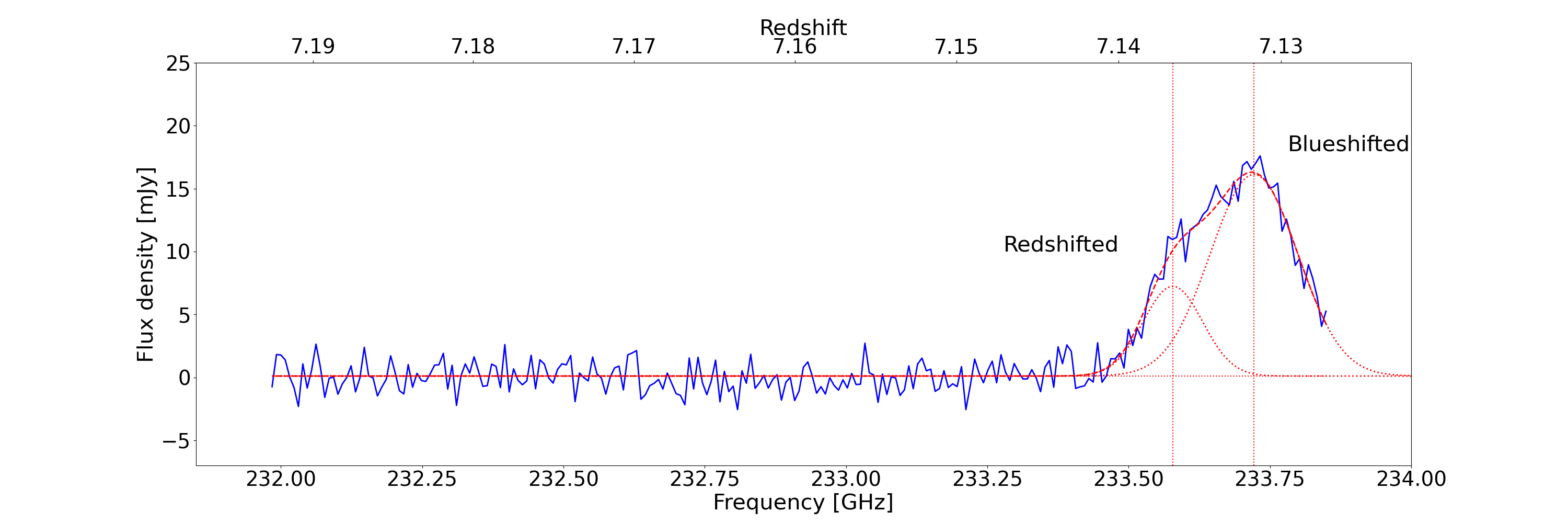}{0.9\linewidth}{(a) [\ion{C}{2}] spectrum}}
    \gridline{\fig{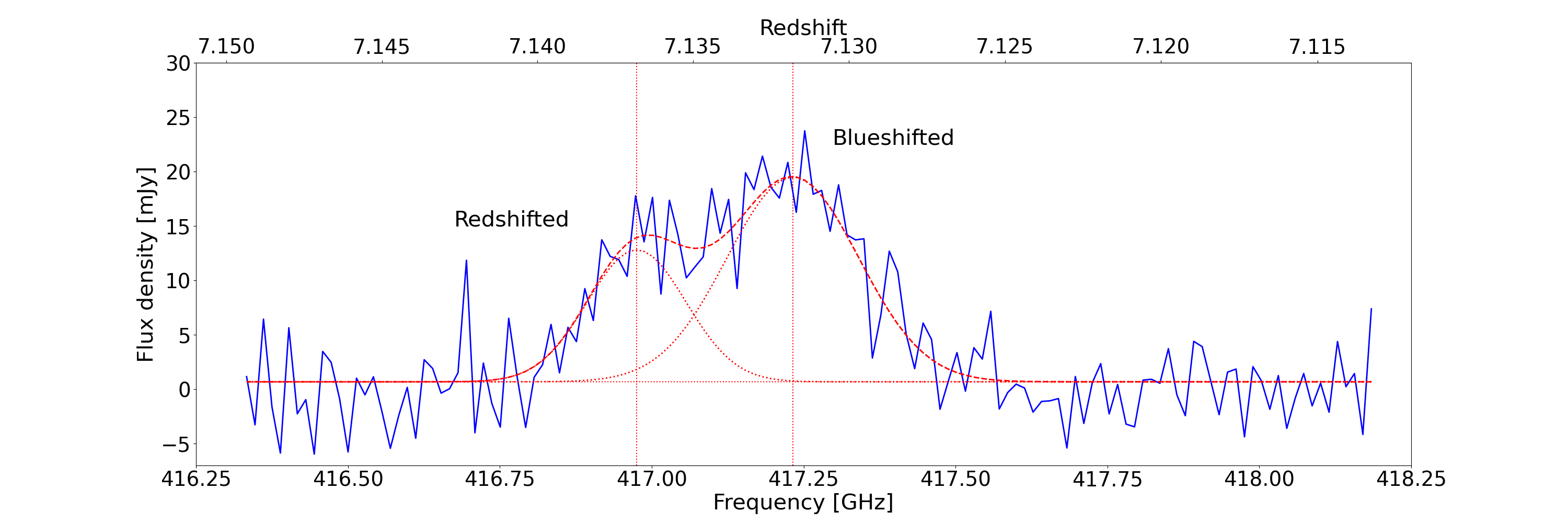}{0.9\linewidth}{(b) [\ion{O}{3}] spectrum}}
    \caption{Spectra of (a) [\ion{C}{2}]158\,{\rm{$\mu$m}} and (b) [\ion{O}{3}]88\,{\rm{$\mu$m}} using ALMA (a) Band 6 and (b) Band 8 data (natural \& $uv$-tapered). The red dotted lines in each panel show the Gaussian fitting results of clumps A and B as indicated.
\label{Fig:spectra}}
\end{figure*}

Using elliptical apertures of 8$^{\prime\prime}$.5$\times$7$^{\prime\prime}$.5 and 6$^{\prime\prime}$.5$\times$5$^{\prime\prime}$.5 (both approximately 3 beams), we extracted spectra from the natural and $uv$-tapered Band 6 and Band 8 image cubes, respectively.
The spectra are shown in blue in Fig.~\ref{Fig:spectra}.
This is to include all the flux obtainable by covering the area where the spatially extended emission may exist, for instance, the [\ion{C}{2}]158\,{\rm{$\mu$m}} halos around $z\sim6$ galaxies reported by \citet{2019ApJ...887..107F}.
At around the rest frequencies of [\ion{C}{2}]158\,{\rm{$\mu$m}} and [\ion{O}{3}]88\,{\rm{$\mu$m}}, in Band 6 and Band 8 respectively, we detect a line emission with two peaks in each band, indicated as `redshifted' and `blueshifted' as shown in Fig.~\ref{Fig:spectra}.
They may indicate two 
merging components.
More analyses are presented in sections~\ref{sec:ratio and luminosity results} and \ref{sec:velocity fields results}.
Here, we first fit the spectra with two Gaussians.
We obtain two peaks of Gaussians centering at 233.58$\pm$0.05\,GHz and 233.72$\pm$0.08\,GHz, for [\ion{C}{2}] using Band 6 data in panel (a).
Similarly, for Band 8 data, we find two peaks of [\ion{O}{3}] at 416.98$\pm$0.08\,GHz and 417.23$\pm$0.11\,GHz, respectively, after applying two-Gaussian fitting as shown in panel (b).
The fitted results are shown as red dotted lines in Fig.~\ref{Fig:spectra}.
We acquire a redshift of $z_{\rm{[C\,II]red}}=7.137\pm0.002$ for the redshifted part, and that of $z_{\rm{[C\,II]blue}}=7.132\pm0.003$ for the blueshifted part, for [\ion{C}{2}] line detections in Band 6.
Accordingly, we obtain redshifts of $z_{\rm{[O\,III]red}}=7.137\pm0.002$ and $z_{\rm{[O\,III]blue}}=7.132\pm0.002$ with Band 8 [\ion{O}{3}] detections, for the redshfited and blueshifted parts.
We find that the redshifts we obtain from the [\ion{C}{2}] and [\ion{O}{3}] line detections are in good agreement with each other for both of the components.

The signal-to-noise ratio (S/N), defined as the ratio of the peak flux of the source (moment-0 map) to the background noise of the image, is calculated for both Band 6 and Band 8 (natural and $uv$-tapered) data, with a value of S/N = =51.5 for [\ion{C}{2}] and S/N = 22.3 for [\ion{O}{3}]. Note that for Briggs weighted data, the S/N values are calculated as 16.1 for [\ion{C}{2}], and 27.9 for [\ion{O}{3}].


Taking the average of the [\ion{O}{3}] and [\ion{C}{2}] redshifts, we obtain a systemic redshift $z_{\rm{sys}} = 7.133\pm$0.005.
This matches the adopted redshift $z=7.13$ in \citet{2021MNRAS.tmpL..91B}. The new redshift derived from FIR fine structure [\ion{C}{2}] and [\ion{O}{3}] lines, and hence is a better measurement than the previous estimate of $z\sim7.5$ based on the Lyman 
break with no detection of 
any emission line \citep{2015Natur.519..327W}.

We measure the widths of the emission lines with their full widths at half maxima (FWHMs) in terms of 
velocity (km\,s$^{-1}$).
Details of the measured and derived parameters, corrected with a lensing magnification factor of $\mu=$9.3 \citep{2008ApJ...678..647B}, of the galaxy are summarized in Table~\ref{Tab:Parameters}.
We also include 1$\sigma$ uncertainty of the parameters in Table~\ref{Tab:Parameters}.
The emission-line fluxes of [\ion{C}{2}] and [\ion{O}{3}] are measured by integrating the flux density along velocity, obtaining a total of 548$\pm$56
\,mJy\,km\,s$^{-1}$ for [\ion{C}{2}], and 640$\pm$69
\,mJy\,km\,s$^{-1}$ for [\ion{O}{3}].
We note that each flux error mentioned above is calculated by a quadrature sum of (i) fitting error of the emission line and (ii) 10\% error due to the uncertainty in the absolute flux scale of ALMA data.
With total line fluxes and systemic redshift deduced, we derive the line luminosities with the equation:
\begin{equation}
    L_{\rm line} = 1.04 \times 10^{-3}\times \left( \frac{S_{\rm line}\,\Delta v}{\rm Jy\,km\,s^{-1}} \right) \left( \frac{D_{\rm L}}{\rm Mpc} \right)^2 \left( \frac{\nu_{\rm obs}}{\rm GHz} \right) \,L_{\odot}
\end{equation}
\citep{1972gcpa.book.....W,1992ApJ...398L..29S,2013ARA&A..51..105C}, where $S_{\rm{line}}\Delta v$ is the measured flux of the line, $D_{\rm{L}}$ is the luminosity distance, and $\nu_{\rm{obs}}$ is the observed frequency.
The total luminosity of [\ion{C}{2}] obtained is (69.2$\pm$1.2
)$\times$10$^{7}$
\,$L_{\odot}$, and that of [\ion{O}{3}] is (144.5$\pm$5.7
)$\times$10$^{7}$
\,$L_{\odot}$.
The obtained $L_{\rm{[C\,II]}}$ is slightly different from the value of (6.1$\pm$0.7)$\times$10$^8$\,$L_{\odot}$ adopted by \citet{2021MNRAS.tmpL..91B} by nearly 1$\sigma$. Possible reasons may include different integration regions, beam sizes of the images, the parameter settings of $uv$-tapering, and/or different frequency ranges included.

\begin{deluxetable*}{lcccc}
\tablenum{3}
\tablecaption{Measured and derived parameters of A1689-zD1, with lensing magnification ($\mu=$9.3) corrected\label{Tab:Parameters}}
\tablewidth{0pt}
\tablehead{
\colhead{} & \multicolumn2c{[\ion{C}{2}]158\,{\rm{$\mu$m}}} & \multicolumn2c{[\ion{O}{3}]88\,{\rm{$\mu$m}}} \\
\colhead{peak} & \colhead{redshifted} & \colhead{blueshifted} & \colhead{redshifted} & \colhead{blueshifted}
}
\startdata
Central frequency (GHz) & 233.58 $\pm$ 0.05 & 233.72 $\pm$ 0.08 & 416.98 $\pm$ 0.08 & 417.23 $\pm$ 0.11 \\
Redshift & 7.137 $\pm$ 0.002 & 7.132 $\pm$ 0.003 & 7.137 $\pm$ 0.002 & 7.132 $\pm$ 0.002 \\
Emission-line width (km s$^{-1}$) & 150 $\pm$ 20 & 237 $\pm$ 19 & 122 $\pm$ 21 & 211 $\pm$ 21 \\
Total emission-line width (km s$^{-1}$) & \multicolumn2c{323 $\pm$ 8} & \multicolumn2c{311 $\pm$ 17} \\
Emission-line flux (mJy km s$^{-1}$) & 124
$\pm$13 & 423
$\pm$43 & 165
$\pm$18 & 475
$\pm$53\\
Total line flux (mJy km s$^{-1}$) & \multicolumn2c{548 
$\pm$ 56} & \multicolumn2c{640 
$\pm$ 69} \\
Luminosity ($10^{7} L_{\odot}$) & 15.7
$\pm$1.6 & 53.5
$\pm$5.5 & 37.2
$\pm$4.2 & 107.3
$\pm$12.0\\
Total luminosity ($10^{7} L_{\odot}$) & \multicolumn2c{69.2 
$\pm$ 7.0} & \multicolumn2c{144.5 
$\pm$ 15.5}
\enddata
\tablecomments{$^{*}$All errors are shown in 1$\sigma$. 
$^{**}$``Emission-line width", ``Emission-line flux", and ``Luminosity" indicate those of separate components derived by the spectral fitting analysis, while ``Total emission-line width", ``Total line flux", and ``Total luminosity" indicate those of the redshifted and blueshifted line emission as a whole. $^{***}$For the errors of fluxes and luminosities, other than the fitting errors that we obtain, we also include 10\% errors in the quadratures due to the absolute calibration uncertainty of ALMA data.}
\end{deluxetable*}

\subsection{[\texorpdfstring{\ion{O}{3}}{OIII}]/[\texorpdfstring{\ion{C}{2}}{CII}] ratio and Luminosity}\label{sec:ratio and luminosity results}

$L_{\rm{[C\,II]}}$ and $L_{\rm{[O\,III]}}$ are calculated using the derived redshift of $z=7.133\pm0.005$, so that 
$L_{\rm{[C\,II]}}=(69.2\pm1.2)\times 10^7 L_{\odot}$ and $L_{\rm{[O\,III]}}=(144.5\pm5.7)\times 10^7 L_{\odot}$. We, thus, obtain a total [\ion{O}{3}]-to-[\ion{C}{2}] luminosity ratio of 
$2.09\pm0.09$.

With the obtained $L_{\rm{[C\,II]}}$, we can also revise the $SFR$
of the system using the relation:
\begin{equation}
    SFR_{\rm{[C\,II]}}=1.0\times10^{-7} \left(\frac{L_{\rm{[C\,II]}}}{L_{\odot}}\right)^{0.98},
\end{equation}
in \citet{2011MNRAS.416.2712D}, which assumed a Kroupa Initial Mass Function (IMF) \citep{2001MNRAS.322..231K}. We obtain a $SFR_{\rm{[C\,II], Krou}} =$ 
46.1$\pm$0.8\,M$_{\odot}$\,$\rm{yr^{-1}}$. In \citet{2021MNRAS.tmpL..91B}, however, their dust-obscured SFR was derived from an IR luminosity-to-SFR relation which assumes a Salpeter IMF, with an estimated value of $SFR_{\rm{obsc,Sal}} =$ 33$\pm$9\, M$_{\odot}\,$yr$^{-1}$. 
In order to compare the SFR value we obtain with that derived in \citet{2021MNRAS.tmpL..91B}, we divide our SFR value by a factor of 0.67 to convert it to a Salpeter IMF based SFR \citep{2014madau}; thus, it becomes 
$SFR_{\rm{[C\,II],Sal}}=$ 
68.8$\pm$1.2\,M$_{\odot}$\,$\rm{yr^{-1}}$.
This value is about 2 times the value derived by \citet{2021MNRAS.tmpL..91B}. 
The exact reason for the difference is not clear. Possible reasons include: (1) that the relation from \citet{2011MNRAS.416.2712D} was derived from photometric data corrected for attenuation, while \citet{2021MNRAS.tmpL..91B} derived the SFR from SED fitting result; (2) temperature dependence of [\ion{C}{2}]-SFR relation \citep{1997ApJ...491L..27M}; and/or (3) uncertainty in the SED fitting due to limited FIR photometric coverage. 
Additional further FIR data points would improve the SED fitting. Despite the difference, according to the bottom panel of Fig. 4 in \citet{2011MNRAS.416.2712D}, the logarithmic difference between the SFR derived from attenuation and that from [\ion{C}{2}] line is still within a reasonable range. \citet{2011MNRAS.416.2712D} presented almost an order of scatter on the SFR-[\ion{C}{2}] relation, along with detailed discussions.

\begin{figure}[ht!]
    \epsscale{1.35}
    \plotone{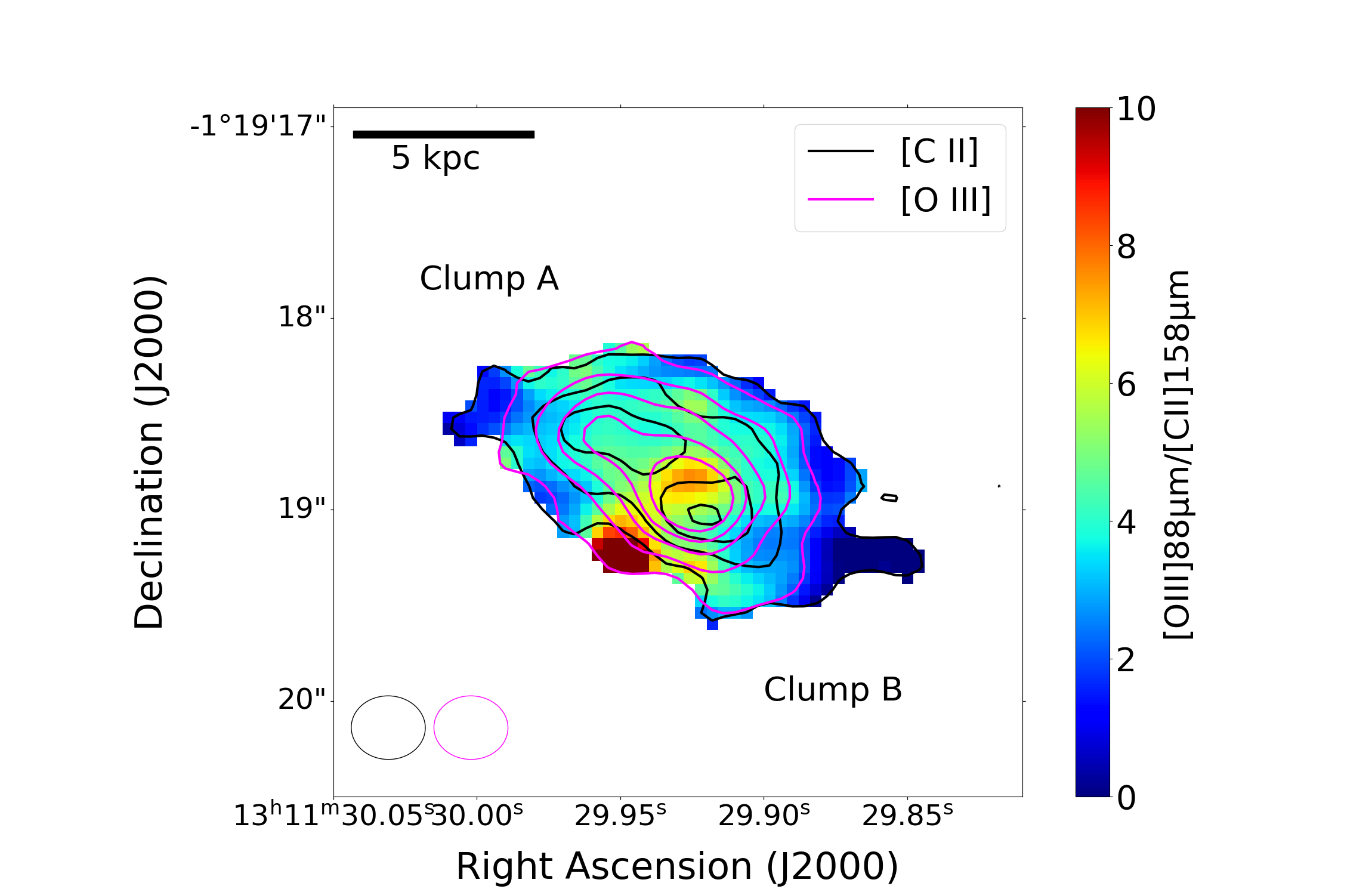}
    \caption{The [\ion{O}{3}]88$\mu$m-to-[\ion{C}{2}]158$\mu$m ratio map. Black contours show [\ion{C}{2}]88$\mu$m line emission, while magenta contours show [\ion{O}{3}] emission. [\ion{C}{2}] contours are shown at 3, 7, 11 and 15$\sigma$ levels, where $\sigma_{\rm{[C\,II]}}\sim$17.1\,mJy\,beam$^{-1}$\,km s$^{-1}$. [\ion{O}{3}] contours are shown at 3, 7, 11, 15, 19$\sigma$ levels, where $\sigma_{\rm{[O\,III]}}\sim$55.7\,mJy\,beam$^{-1}$\,km s$^{-1}$. The color bar shows the value of [\ion{O}{3}]-to-[\ion{C}{2}] luminosity ratio. The smaller beam size ($0^{\prime\prime}.271\times0^{\prime\prime}.243$) and finer pixel size (0$^{\prime\prime}$.035/pix $\times$ 0$^{\prime\prime}$.035/pix) of the corresponding Band 6 (Briggs) are enlarged to ($0^{\prime\prime}.387\times0^{\prime\prime}.332$; 0$^{\prime\prime}$.060/pix $\times$ 0$^{\prime\prime}$.060/pix) 
    to match those of Band 8 data (Briggs). The pixels with [\ion{C}{2}] and [\ion{O}{3}] fluxes lower than 3 $\sigma$ are masked out. The resulting beam sizes are shown on the lower left of the figure.}
    \label{Fig:ratio map}
\end{figure}

We show the [\ion{O}{3}]/[\ion{C}{2}] luminosity ratio map overlaid by black contours of [\ion{C}{2}] 
emission-line flux at 3, 7, 11 and 15$\sigma$ levels, and magenta contours of [\ion{O}{3}], emission-line flux at 3, 7, 11, 15 and 19$\sigma$ levels in Fig.~\ref{Fig:ratio map}.

In order to make a comparison between the two emissions, 
we need to enlarge the beam size of Band 6 [\ion{C}{2}] data to that of Band 8 [\ion{O}{3}] data before taking the ratio. Therefore, we apply a $uv$-tapering in the \texttt{CASA} task \texttt{tclean()} to the [\ion{C}{2}] Briggs image to reach a spatial resolution that roughly matches that of the [\ion{O}{3}] Briggs image, and further match the synthesized beam size and position angle of the [\ion{O}{3}] cube exactly with the \texttt{tclean()} parameter \texttt{restoringbeam}. 
After matching the beam size of [\ion{C}{2}] data to that of [\ion{O}{3}] data, we use a python package \texttt{reproject} to reproject the finer pixel size of [\ion{C}{2}] data from  from (0$^{\prime\prime}$.035/pix $\times$ 0$^{\prime\prime}$.035/pix) to (0$^{\prime\prime}$.060/pix $\times$ 0$^{\prime\prime}$.060/pix), in order to match that of [O III] data (0$^{\prime\prime}$.060/pix $\times$ 0$^{\prime\prime}$.060/pix). Then, we divided the two arrays and re-created the [\ion{O}{3}]/[\ion{C}{2}] ratio map.

This is one of the rare cases that a spatial distribution of [\ion{O}{3}]/[\ion{C}{2}] map is revealed at $z>$7.
Interestingly, [\ion{O}{3}]/[\ion{C}{2}] ratio is the highest around the center of the galaxy with a value of $\sim$7, and gradually decreases toward the outer parts of the galaxy, reaching a ratio of one.
We further discuss the implications in section~\ref{subsec:high luminosity ratio discuss}.

\begin{figure}[ht!]
\epsscale{1.25}
\plotone{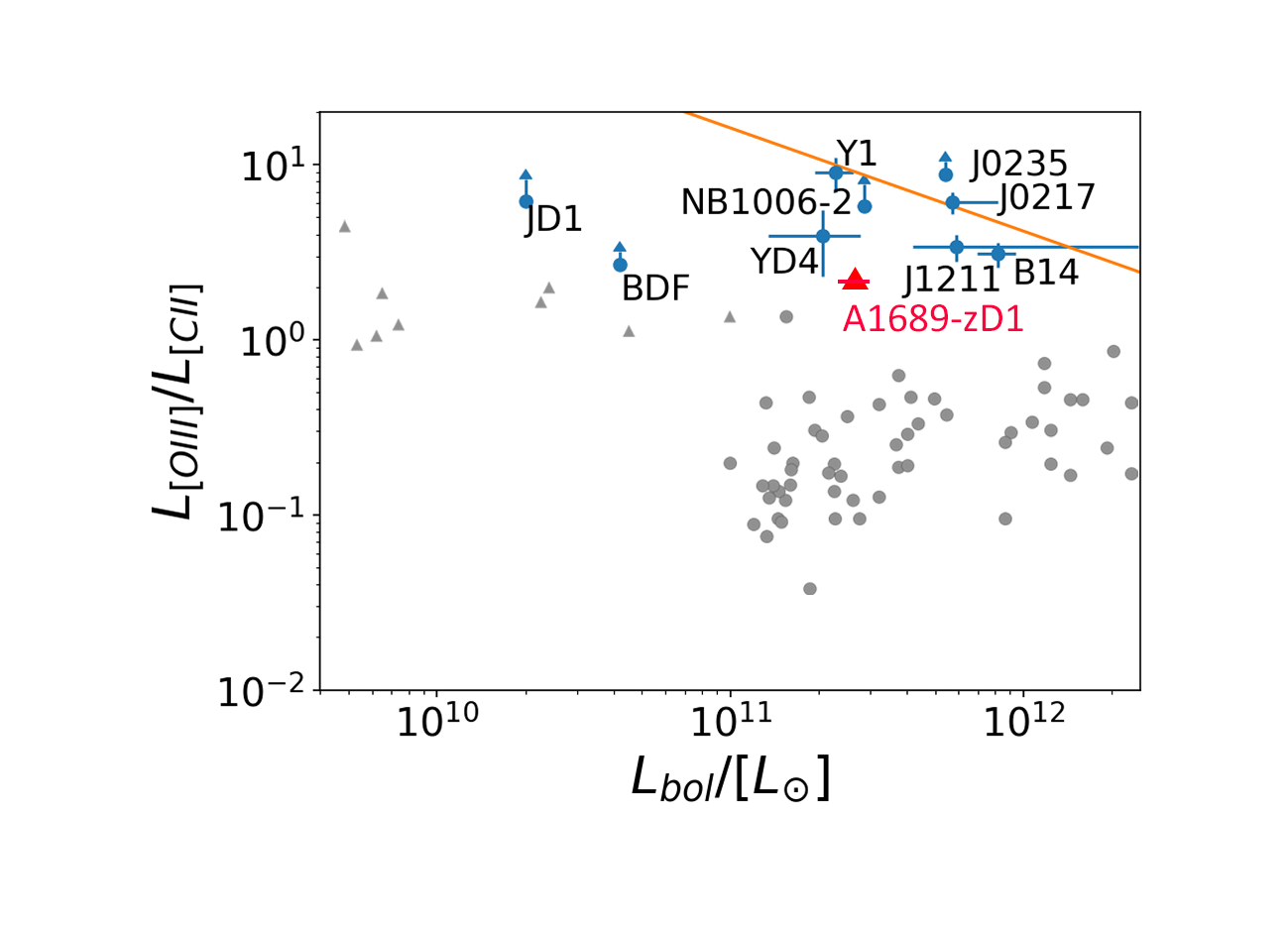}
\caption{[\ion{O}{3}]88$\mu$m/[\ion{C}{2}]158$\mu$m luminosity ratio as a function of bolometric luminosity. The red triangle shows the data point for A1689-zD1. The bolometric luminosities of $z=6-9$ galaxies are taken from \citet{2019PASJ...71...71H}, while the orange line is adopted from Equation (7) in \citet{2020ApJ...896...93H}, which is the fitting function of $z=6-9$ galaxies.
The gray triangle and circles denote $z\sim$0 galaxies from the Dwarf Galaxy Survey \citep{2013PASP..125..600M,2014A&A...568A..62D,2015A&A...578A..53C} and Great Observatories All-sky LIRG Survey \citep[GOALS:][]{2010ApJ...715..572H,2017ApJ...846...32D}, respectively.
 The bolometric luminosity is estimated as a total of UV and TIR luminosities.}
\label{Fig:bolometric-ratio}
\end{figure}

In Fig.~\ref{Fig:bolometric-ratio}, we show the [\ion{O}{3}]/[\ion{C}{2}] luminosity ratio as a function of the bolometric luminosity, which is defined as the summation of the UV and total infrared (TIR) luminosities, i.e., $L_{\rm{bol}} = L_{\rm{UV}} + L_{\rm{TIR}}$ \citep{2019PASJ...71...71H,2020ApJ...896...93H}.
For comparisons, we also plot the ratios for $z = 6$–9 galaxies in the literature plotted in \citet{RN36} before applying Surface Brightness Dimming (SBD), as well as those of local galaxies studied in the Dwarf Galaxy Survey \citep{2013PASP..125..600M,2014A&A...568A..62D,2015A&A...578A..53C} and the Great Observatories All-sky LIRG Survey \citep[GOALS:][]{2010ApJ...715..572H,2017ApJ...846...32D}. 
The orange line is the fitting function of $z=6$-9 galaxies without A1689-zD1, adopted from Equation (7) in \citet{2020ApJ...896...93H}, as a comparison. Although as suggested in \citet{RN36}, SBD may affect the flux measurements gravely with ALMA, we do not apply the same correction because of the high S/N ratio of [\ion{C}{2}] emission of A1689-zD1 ($\sim$17.1). Such an effect was predicted to be weak at high S/N ratio \citep{RN36}.
We find that A1689-zD1 and other $z = 6$–9 galaxies show systematically higher [\ion{O}{3}]/[\ion{C}{2}] ratios than most of the local galaxies, which is consistent with previous results \citep{2016Sci...352.1559I,2019MNRAS.487L..81L}.

\subsection{Velocity fields}\label{sec:velocity fields results}

\begin{figure}[ht!]
\gridline{\fig{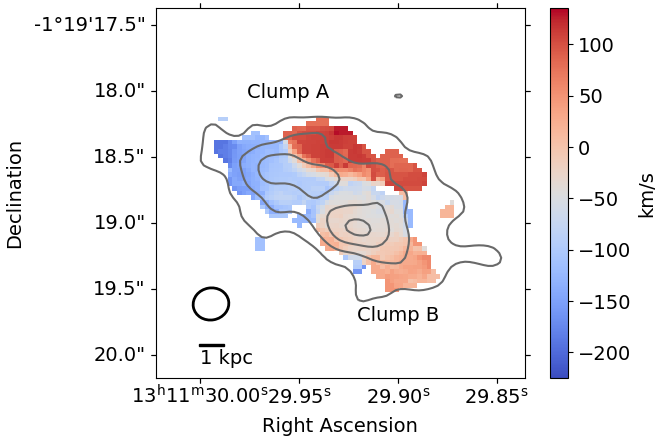}{0.45\textwidth}{(a) [\ion{C}{2}]158\,{\rm{$\mu$m}} velocity field}}
\gridline{\fig{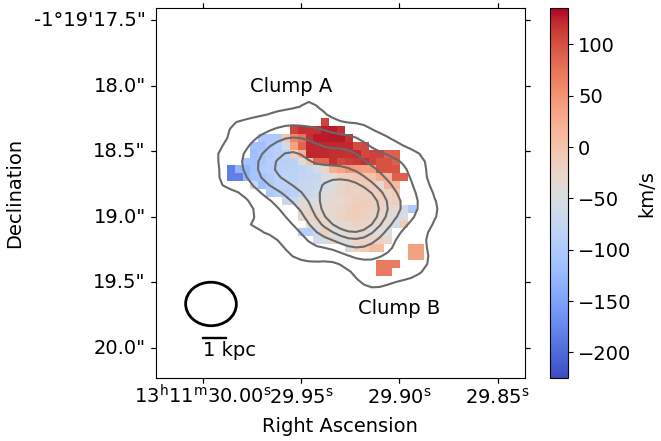}{0.45\textwidth}{(b) [\ion{O}{3}]88\,{\rm{$\mu$m}} velocity field}}
\caption{The velocity field (moment 1 map) of the emission lines (top: [\ion{C}{2}]158$\mu$m and bottom: [\ion{O}{3}]88$\mu$m; using Briggs weighted data). The gray contour in the top (bottom) panel is the same as the white (magenta) contour for [\ion{C}{2}] ([\ion{O}{3}]) emission in Figure.~\ref{Fig:mom0}. Color bar in each panel shows the velocity in the range of -225 km s$^{-1}\sim$135 km s$^{-1}$, as it is close to and covers the maximum and minimum values of both emissions.
\label{Fig:mom1}}
\end{figure}

In this section, we investigate the kinematic properties of A1689-zD1 using [\ion{O}{3}] and [\ion{C}{2}] emissions.
With the \texttt{CASA} task \texttt{immoments}, we create flux-weighted velocity (i.e., moment 1) maps of [\ion{C}{2}] and [\ion{O}{3}] emission lines (Fig.~\ref{Fig:mom1}), using pixels with $>$2$\sigma$ detections of Briggs data. Note that we do not consider the beam smearing effect for the velocity fields.

Comparing the clumps A and B we observe in Fig.~\ref{Fig:mom0}, we see that clump A is consisted of a blueshifted component on the upper left (east side) and a redshifted component on the upper right (west side), while clump B is a cloud of redshifted and blueshifted mixture, but not as intense as the components in clump A.


We only
investigate the 
[\ion{C}{2}] field because:
(1) Band 6 data ([\ion{C}{2}]) has a better S/N
compared to Band 8 data ([\ion{O}{3}]), and
(2) the patterns and the values of velocity field in [\ion{C}{2}] and [\ion{O}{3}] are similar.

A1689-zD1 might be a merger due to the parallel isovelocity lines found between clump A and clump B, as how it has been claimed \citep{2008ApJ...678..647B,2017MNRAS.466..138K}. A comparison to Fig.~\ref{Fig:mom0} shows that the two peaks of [\ion{C}{2}] are consistent with clump B and the blueshifted part of clump A, while the redshifted part of clump A could possibly be (1) some ejected materials from the merger, or (2) from a highly obscured component of the galaxy. Similar discussions can be found in \citet{2017ApJ...850..180J} and \citet{2019PASJ...71...71H}, while further comparisons to other targets are presented in section~\ref{subsec:velocity fields discuss}.



\section{Discussion}\label{sec:discussion}




\subsection{High [\texorpdfstring{\ion{O}{3}}{OIII}]/[\texorpdfstring{\ion{C}{2}}{CII}] luminosity ratio}\label{subsec:high luminosity ratio discuss}

From section~\ref{sec:ratio and luminosity results}, we find a high [\ion{O}{3}]/[\ion{C}{2}] luminosity ratio of $2.17\pm0.14$.
In the literature \citep[e.g.,][]{2020MNRAS.493.4294B,2019MNRAS.487L..81L,2017A&A...605A..42C}, the [\ion{O}{3}]/[\ion{C}{2}] luminosity ratios of the $6<z<9$ galaxies have been found systematically larger than those of $z\sim$0 galaxies, similar to A1689-zD1.
Formerly, \citet{2020ApJ...896...93H} also found high values of the ratio for their large samples of high-$z$ galaxies.

From their simulations in \citet{katz2021nature}, when they assume a low C/O abundance, their core collapse supernova model agrees well with observations with high [\ion{O}{3}]/[\ion{C}{2}] luminosity ratios at high redshift.
Therefore, a low C/O abundance may account for the high [\ion{O}{3}]/[\ion{C}{2}] ratios for $z\sim6$ galaxies.

Moreover, as mentioned in \citet{RN36}, there might also be an underestimation of [\ion{C}{2}] flux caused by high angular resolution.
Often [\ion{C}{2}] emission is more spatially extended, and thus, such extended flux might be underestimated due to the interferometric nature of ALMA, which is why we further perform $uv$-tapering so as to increase the beam size of Band 6 data to $\sim$2$^{\prime\prime}$, with a peak S/N of 17.1.
Comparing our results with those in \citet{RN36}, the value of $L_{[\rm{O\,III]}}/L_{[\rm{C\,II]}}=2.09\pm0.09$ of A1689-zD1 just touched the lower limit of 
the range of $2<L_{\rm{[O\,III]}}/L_{\rm{[C\,II]}}<8$ measured with the nine observed high-$z$ galaxies reported in \citet{RN36}.
The comparatively high values of $L_{[\rm{O\,III]}}/L_{[\rm{C\,II]}}$ for the other $z>6$ sources may possibly be due to observational limitations. \citet{RN36} found that for the nine $z>6$ galaxies, up to 40 percent of the extended [\ion{C}{2}] component might be missed at an ALMA angular resolution of 0$^{\prime\prime}$.8, implying that $L_{[\rm{C\,II]}}$ would possibly be underestimated by a factor of $\sim$2 in data at low S/N ($<$7).

In the models presented in \citet{2020ApJ...896...93H}, they suggested several possible mechanisms responsible for the high ratio of [\ion{O}{3}]/[\ion{C}{2}] luminosities that might match with the case for the galaxy:
(1) higher ionization parameter ($U_{\rm{ion}}$) in higher redshift galaxies, possibly due to the low-metallicity young stellar populations with larger volumes of [\ion{H}{2}] regions;
(2) low photodissociation region (PDR) covering fraction $C_{\rm{PDF}}$, with the definition being:
\begin{equation}
    C_{\rm PDF} = \frac{\rm No.\,of\,sightlines\,with\,PDRs}{\rm No.\,of\,sightlines\,with\,H~II~regions}
\end{equation}
where $C_{\rm{PDF}}=$ 0 gives a decrease of $L_{[\rm{C\,II]}}$ by $\sim99\%$; and
(3) low C/O ratio due to a young stellar age.
Since a higher $U_{\rm{ion}}$ gives rise to a more extended \ion{H}{2}, and thus [\ion{O}{3}], region (Fig.15(b) of \citet{2020ApJ...896...93H}), it may not be able to explain the observed sharp peak of [\ion{O}{3}] at the center of A1689-zD1.
A low $C_{\rm{PDF}}$ leads to a low [\ion{C}{2}] luminosity. While it is impossible that no PDR overlaps with the \ion{H}{2} region ($C_{\rm{PDF}}=0$), a low coverage of PDR region ([\ion{C}{2}] emission) together with a concentrated \ion{H}{2} region may still explain the peaky [\ion{O}{3}] as observed in Fig.~\ref{Fig:ratio map}.
A low C/O ratio due to production of C at young stellar age may partially explain the high [\ion{O}{3}]/[\ion{C}{2}] ratio. But it does not indicate the spatial distribution of the ratio.

From Fig.~\ref{Fig:bolometric-ratio}, most of the discovered high-z galaxies ($z=$6-9) have an overall $L_{[\rm{O\,III}]}/L_{[\rm{C\,II}]}$ higher than that of A1689-zD1 ($\sim$2).
Thus, we expect that other z$\sim$7 galaxies might obtain even higher values of the luminosity ratio [\ion{O}{3}]/[\ion{C}{2}] at the center if they are observed with higher spatial resolution.

Apart from that, in Fig.~\ref{Fig:ratio map}, we find that the [\ion{O}{3}]/[\ion{C}{2}] ratio is the highest in between the optical peaks shown in the HST images (clumps A and B; Fig.~\ref{Fig:mom0}), but not at the peaks, with a ratio of $\sim7$, then gradually decreases outward.
This is a high value as compared to an average luminosity ratio of $L_{\rm{[O\,III]}}/L_{\rm{[C\,II]}}=2.09\pm0.09$.
Thanks to the high spatial resolution of our ALMA data, we have a precious opportunity to investigate the spatial variation of the ratio within the galaxy.

A number of physical mechanisms could cause the high ratio at the center of the galaxy. If a central active galactic nucleus (AGN) is present, the ionization of neutral gas by an AGN may be one reason \citep{2018ApJ...869L..22W}.
Alternatively, if clumps A and B are in the process of merging, the excess [\ion{O}{3}] may also be the heated ionized gas due to shock heating from mergers \citep{2007ApJ...654..731H,2020ApJ...894..157M}. 
Although in such a case, we might expect an extended shock front, rather than a point-like high ratio region we observe in Fig.~\ref{Fig:ratio map}. The velocity difference between clumps A and B is also small, at least in the line of sight direction.
Another explanation may be that A1689-zD1 has a central starburst region, which gives rise to ionized gas \citep{1997ApJ...481..703S}. However, it may not be the case for the sharp peak of [\ion{O}{3}]/[\ion{C}{2}] in A1689-zD1, since, according to \citet{2018A&A...611A..95W}, ionized gas is diffuse in the \ion{H}{2} region of starburst galaxies.

\subsection{Velocity fields}\label{subsec:velocity fields discuss}
The complexity of the velocity field of the galaxy A1689-zD1 shown in Fig.~\ref{Fig:mom1} gives little hint of any existence of a rotating disk. We, therefore, suggest that it may simply be a merging system, in agreement with previous analyses \citep[e.g.,][]{2008ApJ...678..647B,2017MNRAS.466..138K}.


Considering the observed high-$z$ quasars (QSOs) in the literature, in \citet{2019ApJ...881L..23B}, for instance, a host galaxy of a QSO ULAS J1342+0928 at $z=7.54$ showed a velocity gradient with the northern part blueshifted and the southern part redshifted.
In spite of that, the velocity dispersion did not resemble a coherent rotating structure.
Hence, they interpreted the velocity dispersion structure as that of a merger.
On the other hand, \citet{2017ApJ...845..138S} found that the velocity dispersion of a QSO at $z=6.13$, with a best-fit inclination angle of 34$^{\circ}$, has a coherent rotating structure being consistent with the rotation of the galaxy.
Similarly, \citet{2013ApJ...773...44W} observed 6 QSOs at $z\sim6$ using ALMA, and find that their velocity gradients are consistent with rotating, gravitationally bounded gas components.

For [\ion{C}{2}] emitters, \citet{2018Natur.553..178S} found 
rotating structures based on simulation and their velocity fields in COS-3018555981 and COS-2987030247 at $z=6.8$.
Their results are consistent with rotationally supported galaxy disks, which were often found at $z\sim$2.
Likewise, \citet{2020MNRAS.493.4294B} found a Lyman-break galaxy (LBG) (MACS0416-Y1) at $z\sim8.3$ which shows 
a rotation-dominated disk given the observed velocity gradient.
In addition, \citet{2019PASJ...71...71H} used ALMA and detected two clumps in B14-65666 at $z=7.15$ with [\ion{O}{3}] and [\ion{C}{2}] emission lines. However, they believe the galaxy is a starburst galaxy induced by a major merger because they did not see a smooth velocity field.

Given the complex velocity field of A1689-zD1, and the possibility that other dynamical interpretations would be equally valid as supported by other literatures, further investigation is needed to make a conclusion. As mentioned in section~\ref{sec:velocity fields results}, A1689-zD1 may possibly be a merger with some northwestern ejected materials, or a merger with northwestern redshifted materials coming from a third, more highly obscured region of the galaxy.


\section{Conclusions}\label{sec:conclusions}
Using the new ALMA Bands 6 and 8 data, we detect [\ion{C}{2}]158$\mu$m and [\ion{O}{3}]88$\mu$m emission lines 
for A1689-zD1. Our findings are summarized as follows.

\begin{enumerate}
\item[1.] We measure the redshift of A1689-zD1 as $z=7.133\pm0.005$ based on [\ion{C}{2}] and [\ion{O}{3}] emission lines detected in Band 6 and Band 8 (natural \& $uv$-tapered), respectively. The redshift is consistent with that adopted by \citet{2021MNRAS.tmpL..91B}.
\item[2.] Using the derived $L_{\rm{[C\,II]}}$, 
we estimate a star-formation rate of $SFR_{\rm{[C\,II],Krou}}$
 = 46.1$\pm$0.8 M$_{\odot}$\,$\rm{yr^{-1}}$. Converting it from Kroupa IMF to Salpeter IMF, the value becomes $SFR_{\rm{[C\,II],Sal}}$ = 68.8$\pm$1.2\,M$_{\odot}\,$yr$^{-1}$, which is around a factor of 2 larger than the SED-fitting estimate $SFR_{\rm{obsc,Sal}} =$ 33$\pm$9\,M$_{\odot}\,$yr$^{-1}$ in \citet{2021MNRAS.tmpL..91B}. However, the difference is still within a reasonable range according to Fig.~4 of \citet{2011MNRAS.416.2712D}.
\item[3.] The [\ion{O}{3}]/[\ion{C}{2}] luminosity ratio of this galaxy is $2.09\pm0.09$, which is similar to other high-$z$ galaxies, but much larger than its local counterparts.
\item[4.] Despite a number of average $L_{\rm{[O\,III]}}/L_{\rm{[C\,II]}}\sim$ 2, A1689-zD1 has an exceptionally high spatial ratio of $L_{\rm{[O\,III]}}/L_{\rm{[C\,II]}}\sim$ 7 at the center of the galaxy. This is because [\ion{C}{2}] emission is significantly spatially-extended, while [\ion{O}{3}] is compact at the center. Possible reasons of the compactness of [\ion{O}{3}] includes, a central AGN, shock heating, and/or starburst.
\item[5.] The moment 1 maps of the ALMA Bands 6 and 8 data (Fig.~\ref{Fig:mom1}) show complex velocity fields of A1689-zD1, with northeastern part being blueshifted, southwestern part being redshifted, and an additional northwestern part being comparatively more heavily redshifted. It is, hence, suggested that the galaxy may be a merger with the northwestern redshifted part being: (1) some ejected materials from the merger, or (2) some components coming from a third, more highly obscured region of the galaxy.
\end{enumerate}


\begin{acknowledgments}

We are grateful to the anonymous referee for all the insightful comments. This paper makes use of the following ALMA data: ADS/JAO.ALMA\#2015.1.01406.S and ADS/JAO.ALMA\#2017.1.00775.S. ALMA is a partnership of ESO (representing its member states), NSF (USA) and NINS (Japan), together with NRC (Canada), MOST and ASIAA (Taiwan), and KASI (Republic of Korea), in cooperation with the Republic of Chile. The Joint ALMA Observatory is operated by ESO, AUI/NRAO and NAOJ.
TH is supported by the Centre for Informatics and Computation in Astronomy (CICA) at National Tsing Hua University (NTHU) through a grant from the Ministry of Education of the Republic of China (Taiwan).
TG and TH acknowledge the supports by the Ministry of Science and Technology of Taiwan through grants 108-2628-M-007-004-MY and 110-2112-M-005-013-MY3, respectively.
This research has made use of NASA’s Astrophysics Data System.
\end{acknowledgments}

%

\vspace{5mm}
\facilities{HST, ALMA}


\software{CASA \citep{2007ASPC..376..127M}, astropy \citep{2013A&A...558A..33A,2018AJ....156..123A}
          }




\bibliography{bibliography.bib}{}
\bibliographystyle{aasjournal}



\end{document}